\documentstyle[aps]{revtex}
\newcommand{\nJ}{{\bf J}}
\newcommand{\nr}{{\bf r}}

\title{Nuclear currents based on the integral
form \\of the continuity equation}

\author{P.J. Barneo, J.E. Amaro and A.M. Lallena}

\address{Departamento de F\'{\i}sica
Moderna, Universidad de Granada,
E-18071 Granada, Spain}

\begin{document}
\maketitle

\begin{abstract}
We present an approach to obtain new forms of the nuclear
electromagnetic current, which is based on an integral form of the
continuity equation. The procedure can be used to restore current
conservation in model calculations in which the continuity equation is
not verified. Besides, it provides, as a particular result, the
so-called Siegert's form of the nuclear current, first obtained by
Friar and Fallieros by extending Siegert's theorem to arbitrary values
of the momentum transfer. The new currents are explicitly conserved
and permit a straightforward analysis of their behavior at both low
and high momentum transfers.  The results are illustrated with a
simple nuclear model which includes a harmonic oscillator mean
potential.

\vspace{.5cm}

\noindent
{PACS numbers: 25.30.Dh,23.20.Js }

\end{abstract}


\section{Introduction}

One of the oldest and still unsolved problems in nuclear physics
is that of the determination of the electromagnetic current
operator. The basic difficulty lies on the absence of useful
constraints one can impose to this operator. For a long time,
Siegert theorem \cite{Sie37} has played an important role in
this respect, mainly because its application allows to avoid
the consideration of two-body currents in the calculations.
After the application of the theorem, the
effects of such currents can be evaluated in terms of the charge
contributions, the corrections of which are {\it a priori}\/ small.
However, Siegert theorem only applies in the long-wavelength
limit and then it is irrelevant for the values of the momentum
transfers usually observed in electron scattering experiments.

In principle, one can think that current conservation, and more
precisely the continuity equation (CE), could provide the needed
procedure. Nevertheless, the task cannot be accomplished due to the
impossibility to fix in a unique manner the operator, because those 
terms given as the rotational of any function are not constrained by
the CE. 

Using current conservation, the longitudinal current is
eliminated in terms of the charge operator, and the electron
scattering cross section reads
\begin{equation}
\frac{d\sigma}{d\Omega} =
\frac{4\pi\sigma_M}{f_{rec}(2J_i+1)}
     \left[ v_L\sum_\lambda|t_{C\lambda}|^2
           +v_T\sum_\lambda\left(|t_{E\lambda}|^2
                                +|t_{M\lambda}|^2
                           \right)
     \right]
\end{equation}
where $t_{C\lambda}$, $t_{E\lambda}$, and $t_{M\lambda}$
are the Coulomb, electric and magnetic multipoles of the transition.
It is common to rewrite the matrix elements of the electric
transverse operators in terms of the charge matrix elements, by using
the equations 
\begin{eqnarray}
\frac{\omega}{q}t_{C\lambda} 
&=& -\sqrt{\frac{\lambda}{2\lambda+1}} t_{\lambda-}
    -\sqrt{\frac{\lambda+1}{2\lambda+1}} t_{\lambda+}
\label{tclambda}\\
t_{E\lambda}
&=& \sqrt{\frac{\lambda+1}{2\lambda+1}} t_{\lambda-}
    -\sqrt{\frac{\lambda}{2\lambda+1}} t_{\lambda+}.
\label{telambda}
\end{eqnarray}
Here we have defined the transition matrix elements
\begin{equation}
t_{\lambda\pm} = \langle J_f\| iT_{\lambda,\lambda\pm 1}\| J_i\rangle
\end{equation}
of the multipole operators 
\begin{equation}
\label{multipoles}
T_{JLM}(q) \, = \,
\int {\rm d}^3r\, j_L(qr) {\bf Y}_{JL}^{M}(\hat{\bf r})
\cdot{\bf J}({\bf r}) \, ,
\end{equation}
where $\nJ(\nr)$ is the electromagnetic current operator 
of the nucleus and ${\bf Y}_{JL}^{M}(\hat{\bf r})$ labels a vector spherical
harmonic. 

Eq. (\ref{tclambda}) is the multipole expression of the CE. It
relates the charge multipoles to the longitudinal multipoles of the
current.
Eq. (\ref{telambda}) is the definition of the electric multipoles. 
Using eq. (\ref{tclambda}) it is possible to eliminate one of the
multipoles
$t_{\lambda\pm}$ in terms of the Coulomb $t_{C\lambda}$ multipole
and substitute it in the expression of $t_{E\lambda}$. In this way
one can use the CE constraint even in the calculation
of the transverse form factor. 
Besides, this procedure allows to minimize the errors coming from an 
insufficient knowledge of the transverse current 
(e.g. meson-exchange currents) 
in electron scattering calculations.

This method of rewriting the matrix elements of the electric
transverse operator in terms of the charge matrix elements,
has been applied traditionally
in many electron scattering calculations. Despite
this ensures that CE is satisfied, the procedure is completely
meaningless in models which do not satisfy the CE \cite{Ama96}. 
As a consequence, the problem of current
conservation in model calculations of electron scattering by nuclei is
still under theoretical study \cite{Cap97}-\cite{Nau97}.

A similar situation appears in the relativistic treatment of the
off-shell $\gamma NN$ vertex in (e,e'p) reactions. Several
prescriptions for the current operators obtained by extrapolating the
on-shell currents have been proposed \cite{deF83}. The corresponding 
off-shell vertex operators violate current conservation, which is
enforced by elliminating the 3- or 0-components using the continuity
equation. The problem which arises is that different results are
obtained depending on the option chosen \cite{Cab93}.

An alternative approach is provided by the so-called Siegert current
(SC) developed by Friar and Fallieros \cite{Fri82,Fri84} by extending
the Siegert's theorem to arbitrary wavelengths. This procedure is
based on isolating the components of the current which are constrained
by the CE and replacing them by an adequate combination of the charge
multipoles. However, the results are not satisfactory \cite{Fri85}
because they show a pathological behavior for high momentum transfer
when the nuclear model considered does not verify current conservation.
This has been also pointed out recently by Caparelly and de Passos 
\cite{Cap97} in RPA and TDA calculations. 

Two main conclusions of the work of Friar and Fallieros \cite{Fri84}
must be pointed out. First, the SC appears to be preferable to the 
traditional forms of the nuclear currents even for model which satisfy
the CE, because the size of the exchange current operators
is considerably reduced. Second, they suggest that well behaved current 
operators, satisfying the constraints of current conservation
for high momentum transfer, could be determined, but 
these new forms of the current have not been found up to know.

In this paper we present a new approach to obtain nuclear conserved
currents. Specifically we show that the SC appears as a particular
case of a more general family of current operators which can be
obtained from the CE. Though all of them produce identical results in
models conserving the current, they permit to formulate different
prescriptions for restoring current conservation in electron
scattering model calculations. Besides, our formalism allows us to
obtain explicitly the asymptotic behavior of these new currents.

Our method deals with the CE written in integral form (Sec. II) from
which the new currents can be obtained directly both in coordinate and
momentum spaces. Our approach generalizes the Friar and Fallieros
procedure and, as a consequence, permits to find, {\it a priori},
infinite ways of restoring the CE in models which do not conserve the
current. In Sec. III we show how to do it and we study in detail the
low and high momentum transfer behavior of the conserved currents
found as well as that of their multipoles. To finish, in Sec. IV we
show some examples in a simple nuclear model. The conclusions are
drawn in Sec. V.


\section{Integral continuity equation}


\subsection{Coordinate space}
\label{coord}

Usually, the CE is written in differential form as 
\begin{equation}
\label{conteq}
\nabla \cdot {\bf J}({\bf r}) \, = \, -i\, 
\left[ H,\rho({\bf r}) \right] \, ,
\end{equation}
where $H$ is the Hamiltonian and $\rho$ and ${\bf J}$ are the
nuclear charge and current densities, respectively.
This equation restricts the value of $\nabla\cdot{\bf J}$, once the
density $\rho$ and the Hamiltonian $H$ of the system are known, but,
as mentioned above, it does not permit to fix completely the current
because the rotational terms do not contribute to the divergence.

The key point in our development consists in writing an integral
equation equivalent to the CE. To do that, we start with the 
relation
\begin{equation}\label{e2}
\frac{1}{r}\frac{\rm d}{{\rm d} r}
\left[ r^2 {\bf J}({\bf r}) \right] \, = \, 
2 {\bf J}({\bf r})
+({\bf r}\cdot\nabla){\bf J}({\bf r}) \, .
\end{equation}
Taking into account the vector identity
\[
\nabla\times \left[{\bf r}\times{\bf J}({\bf r})\right]\, = \, 
\left[\nabla\cdot{\bf J}({\bf r})\right]{\bf r}-2{\bf J}({\bf r})-
{\bf r}\cdot\nabla {\bf J}({\bf r}) \, ,
\]
one can rewrite Eq.~(\ref{e2}) as
\begin{equation}\label{e4}
\frac{\rm d}{{\rm d}r}
 \left[ r^2{\bf J}({\bf r}) \right] \, = \, r  \left\{
\left[\nabla\cdot{\bf J}({\bf r})\right] {\bf r}-
\nabla\times \left[{\bf r}\times{\bf J}({\bf r})\right] 
\right\} \, .
\end{equation}
Integrating in the radial direction and using the CE (\ref{conteq}) 
we obtain
\begin{eqnarray} 
r^2{\bf J}({\bf r})-r_0^2{\bf J}({\bf r}_0) &= &\nonumber
\int_{r_0}^{r} {\rm d}r \, r \, 
\left\{  \left[ \nabla\cdot{\bf J}({\bf r}) \right]{\bf r}-
\nabla\times \left[
{\bf r}\times{\bf J}({\bf r}) \right] \right\} \\
&= & \label{e6}
-\int_{r_0}^{r} {\rm d}r \, r \, 
\left\{  i[H,\rho({\bf r})]{\bf r}+
\nabla\times \left[ {\bf r}\times{\bf J}({\bf r})\right] \right\}
\, .
\end{eqnarray}
This equation permits to obtain the integral continuity equation
(ICE)
\begin{equation}\label{ICE}
{\bf J}({\bf r}) \, = \, \frac{r_0^2}{r^2}{\bf J}({\bf r}_0) 
-\frac{1}{r^2} \, \int_{r_0}^r {\rm d}r\, r \, 
  \left\{  i[H,\rho({\bf r})]{\bf r}
  +\nabla\times \left[ {\bf r}\times{\bf J}({\bf r})\right]
  \right\} \, .
\end{equation}

The ICE is equivalent to the CE and if current conservation 
is fulfilled, the electromagnetic current is recovered by computing
the right-hand side of the equation. The ICE is simplified by choosing
a value of ${\bf r}_0$ for which $r_0^2{\bf J}({\bf r}_0)=0$, such as,
for instance, $r_0=0$ or $\infty$. In addition, the fact that the above
expression of the current is manifestly conserved, permits to use it as
a prescription to restore the CE. This point will be 
discussed in the next section.

It is  useful to write the ICE in an equivalent form. One starts from
the equation 
\begin{equation}\label{e7}
\frac{\rm d}{{\rm d}\alpha}
\left[\alpha^2{\bf J}(\alpha{\bf r})\right] \,=\,
2\alpha{\bf J}(\alpha{\bf r})+\alpha^2{\bf r}\cdot \nabla{\bf J}
(\alpha{\bf r})\, =\, \alpha{\bf C}(\alpha{\bf r}) \, ,
\end{equation}
where we have introduced the adimensional parameter $\alpha$ and 
defined the auxiliary current ${\bf C}$. It is easy to check that
\begin{equation}\label{e8}
{\bf C}({\bf r})\equiv -i[H,\rho({\bf r})]{\bf r}
-\nabla \times \left[ {\bf r}\times{\bf J}({\bf r})\right] \, .
\end{equation}
By integrating Eq.~(\ref{e7}) from $\alpha=\alpha_0$ to $\alpha=1$ we
obtain 
\begin{equation}
\label{e9a} 
{\bf J}({\bf r}) \, = \, \alpha_0^2 {\bf J}(\alpha_0{\bf r}) +
\int_{\alpha_0}^1 {\rm d}\alpha\,\alpha{\bf C}(\alpha{\bf r}) \,
\end{equation}
and introducing the form of ${\bf C}$ given by Eq.~(\ref{e8}) we have
\begin{equation}
\label{e9b}
{\bf J}({\bf r}) \, = \, 
\alpha_0^2 {\bf J}(\alpha_0{\bf r}) -
i \left[ H, \int_{\alpha_0}^1 {\rm d}\alpha \alpha^2 
\rho(\alpha {\bf r}) \right]{\bf r}
-\nabla\times \left[ {\bf r}\times
\int_{\alpha_0}^1 {\rm d}\alpha \alpha{\bf J}(\alpha{\bf r}) 
\right] \, .
\end{equation}
As for Eq.~(\ref{ICE}),
this equation simplifies by selecting a value of $\alpha_0$ such as
$\alpha_0^2 {\bf J}(\alpha_0 {\bf r})=0$. It is of particular interest
to choose $\alpha_0=\infty$. In this case we can write the current as
\begin{equation}
{\bf J}({\bf r})= {\bf J}_c({\bf r})+{\bf J}_m({\bf r}),
\end{equation}
where 
\begin{eqnarray}
{\bf J}_c({\bf r}) 
&\equiv & i\left[H,\int_{1}^\infty d\alpha\, \alpha^2\rho(\alpha{\bf r}) 
         \right]{\bf r}
= i\left[H,\int_{0}^1 \frac{d\lambda}{\lambda^4}\,\rho({\bf r}/\lambda) 
         \right]{\bf r}
\label{e12}\\
{\bf J}_m({\bf r})
&\equiv&
\nabla\times\left({\bf r}\times
                  \int_1^\infty \alpha\,{\bf J}(\alpha{\bf r})
            \right)
= \nabla\times\left({\bf r}\times
                  \int_0^1 \frac{d\lambda}{\lambda^3}{\bf J}({\bf r}/\lambda)
            \right)
\label{e13}
\end{eqnarray}
These currents coincide with those obtained by Friar and Fallieros
(see equations (8a)-(8d) and (4c) in Ref. \cite{Fri84}). Then,
Eq.~(\ref{e9b}) can be considered as an extension of the Siegert's
current formulated by these authors. Other values of the parameter 
$\alpha_0$ fulfilling the relation 
$\alpha_0^2 {\bf J}(\alpha_0 {\bf r})=0$ (such as e.g. $\alpha_0=0$)
provide new currents potentially useful to restore the CE in
model calculations. On the other hand, the procedure we have followed
is considerably simpler.
 
To finish this discussion, we remember that if CE is satisfied, the
different currents one can obtain from Eq.~(\ref{e9b}) coincide. Then,
for the two particular cases $\alpha_0=0$ and $\infty$, we have
\begin{equation}
{\bf J}({\bf r}) = \int_{0}^1 d\alpha\,\alpha{\bf C}(\alpha{\bf r})
         = \int_{\infty}^1 d\alpha\,\alpha{\bf C}(\alpha{\bf r}) \, ,
\end{equation}
and, as a result,
\begin{equation}\label{e15}
\int_{0}^\infty d\alpha\,\alpha{\bf C}(\alpha{\bf r})=0
\end{equation}
This is a global consequence of the CE and is
verified by any conserved current. 

\subsection{Momentum space}
\label{momen}

In order to study the behavior of the currents as a function of the
momentum transfer {\bf q}, it is useful to see how the equations
obtained in the previous subsection read in momentum space. The
corresponding version of the ICE can be obtained by calculating the
Fourier transform in Eq.~(\ref{ICE}). However we derive it again to
illustrate the differences between coordinate and momentum spaces in
what refers to the procedure followed.

We start with the equation analogous to Eq.~(\ref{e2}):
\begin{equation}
\label{e2m}
\frac{\rm d}{{\rm d} q}[q{\bf J}({\bf q})] \, = \, 
{\bf J}({\bf q}) + 
{\bf q}\cdot\nabla_{\bf q}{\bf J}({\bf q}) \, .
\end{equation}
If we consider the vector relation:
\[
\nabla_{\bf q} \left[{\bf q}\cdot{\bf J}({\bf q})\right] \, = \,
{\bf J}({\bf q}) + {\bf q}\cdot\nabla_{\bf q}{\bf J}({\bf q}) +
{\bf q}\times\nabla_{\bf q}\times{\bf J}({\bf q}) \, ,
\]
we obtain from Eq.~(\ref{e2m})
\[
\frac{\rm d}{{\rm d} q}[q{\bf J}({\bf q})] \, = \,
\nabla_{\bf q} \left[{\bf q}\cdot{\bf J}({\bf q})\right] -
{\bf q}\times\nabla_{\bf q}\times{\bf J}({\bf q}) \, ,
\]
and, by inserting here the CE in momentum space
\begin{equation}
\label{cem}
{\bf q}\cdot{\bf J}({\bf q}) = [H,\rho({\bf q})] \, ,
\end{equation}
we have
\begin{equation}
\label{e8m}
\frac{\rm d}{{\rm d} q}[q{\bf J}({\bf q})] \, = \,
[H,\nabla_{\bf q} \rho({\bf q})] -
{\bf q}\times\nabla_{\bf q}\times{\bf J}({\bf q}) \, 
\equiv \, -{\bf C}({\bf q}) \, . 
\end{equation}
Here we have introduced the auxiliary current ${\bf C}({\bf q})$
which, as it is easy to check, is the Fourier transform of the current
${\bf C}({\bf r})$ defined in Eq.~(\ref{e8}). The ICE in momentum
space is then obtained by integrating Eq.~(\ref{e8m})
\begin{eqnarray}
\label{ICE00}
{\bf J}({\bf q}) & = &
\frac{q_0}{q}{\bf J}({\bf q}_0) -
\frac{1}{q}\int_{q_0}^{q} {\rm d}q\,{\bf C}({\bf q}) \\
&=& \frac{q_0}{q}{\bf J}({\bf q}_0) +
\frac{1}{q}\int_{q_0}^{q} {\rm d}q\,
\left\{ [H,\nabla_{\bf q}\rho({\bf q})]-{\bf q}
\times\nabla_{\bf q}\times{\bf J}({\bf q})
\right\} \, .
\label{ICEm}
\end{eqnarray}
As in the case of coordinate space, this ICE reduces by choosing a
value $q_0$ for which $q_0{\bf J}({\bf q}_0)=0$.

We can also obtain the ICE in momentum space equivalent to
Eq.~(\ref{e9b}). To do this we introduce an
adimensional parameter $\lambda$ such that
\begin{equation}
\frac{\partial}{\partial\lambda}\left[
\lambda{\bf J}(\lambda {\bf q}) \right] \, = \, 
-{\bf C}(\lambda {\bf q}) \, .
\end{equation}
Integrating this equation in the interval $[\lambda_0,1]$, and
taking into account the definition of the current ${\bf C}$ given in
Eq.~(\ref{e8m}), we obtain for the ICE in momentum space:
\begin{eqnarray}
\label{ICEgen}
{\bf J}({\bf q}) & = & \lambda_0{\bf J}(\lambda_0 {\bf q}) \, -
\, \int_{\lambda_0}^1 {\rm d}\lambda\, {\bf C}(\lambda {\bf q}) \\
\label{ICEm2}
&=& \lambda_0{\bf J}(\lambda_0 {\bf q}) +
\left[ H, \int_{\lambda_0}^1 {\rm d}\lambda\, 
(\nabla_{\bf q}\rho)(\lambda {\bf q}) \right]
-{\bf q}\times\int_{\lambda_0}^1 {\rm d}\lambda\,\lambda 
     (\nabla_{\bf q}\times{\bf J})(\lambda {\bf q}) \, .
\end{eqnarray}

It is worth to note that the second and third terms in this equation
coincide with the Fourier transforms of the ${\bf J}_c$ and ${\bf
J}_m$ currents defined in Eqs.~(\ref{e12}) and (\ref{e13}). Besides,
it can be shown that, for $\lambda_0=0$, the corresponding currents
introduced by Friar and Fallieros \cite{Fri84} are recovered.

Finally, the condition analogous to Eq.~(\ref{e15}) is found by
considering the values $\lambda_0=0$ and $\infty$ in the ICE
(\ref{ICEm2}):
\begin{equation}\label{e30}
\int_0^{\infty} {\rm d}\lambda\, {\bf C}(\lambda {\bf q}) \,= \, 0 \, .
\end{equation}
The integral in this equation allows us to analyze the asymptotic 
behavior of the
currents in momentum space in case the CE is not fulfilled. We discuss
this point in the next section.

\subsection{Multipoles of the current}

In this subsection we obtain the corresponding expressions for the 
multipoles of the current. As it is known \cite{deF66}, 
the electric and magnetic 
multipoles are linear combinations of the multipole operators 
defined in eq. (\ref{multipoles}).
Taking into account Eq.~(\ref{e9a}) we have
\begin{equation}
T_{JLM}(q) \, = \,
\int {\rm d}^3r\, j_L(qr) {\bf Y}_{JL}^{M}(\hat{\bf r})\cdot
\left[ \alpha_0^2 {\bf J}(\alpha_0 {\bf r}) \, + \,
\int_{\alpha_0}^1 d\alpha\,\alpha {\bf C}(\alpha{\bf r}) \right]\, .
\end{equation}
An easy calculation permits to write
\begin{equation}
\label{e30bis}
T_{JLM}(q)\, = 
\, \frac{1}{\alpha_0}\,
  T_{JLM}\left(\frac{q}{\alpha_0}\right) \, + \,
\int_{\alpha_0}^{1} {\rm d}\alpha \, \frac{1}{\alpha^2}\,
C_{JLM}\left(\frac{q}{\alpha}\right) \, ,
\end{equation}
where we have defined the multipoles of the auxiliary current {\bf C}
as
\begin{equation}
C_{JLM}(q)\equiv
\int {\rm d}^3r\, j_L(qr){\bf Y}_{JL}^{M}(\hat{\bf r})
\cdot{\bf C}({\bf r}) \, .
\end{equation}
Finally, writing the integral (\ref{e30bis}) in terms of the variable
$\lambda=1/\alpha$ we obtain
\begin{equation}
\label{multi}
T_{JLM}(q) \, = \, \lambda_0\,T_{JLM}(\lambda_0 q) \, + \,
\int_{1}^{\lambda_0} {\rm d}\lambda \, C_{JLM}(\lambda q) \, ,
\end{equation}
where $\lambda_0=1/\alpha_0$. Hence, the multipoles of the current 
${\bf J}$ are given by integrating the multipoles $C_{JLM}(q)$ 
of the auxiliary current ${\bf C}({\bf r})$. This equation is the one
corresponding to Eq.~(\ref{ICEgen}) verified by the current 
${\bf J}({\bf q})$, and which is, therefore, satisfied by each one
of its multipoles separately. 

Introducing in Eq.~(\ref{multi}) the variable $q^\prime = \lambda q$
we have finally
\begin{equation}\label{e33}
T_{JLM}(q) \, = \, \frac{q_0}{q}T_{JLM}(q_0) \, + \,
 \frac{1}{q}\int_{q}^{q_0} {\rm d}q'\, C_{JLM}(q') \, ,
\end{equation}
where we have defined $q_0=\lambda_0 q$. 

By choosing the values $q_0=0$ and $\infty$, for which the first term
in Eq.~(\ref{e33}) vanishes, we obtain a global condition for the 
integral of the multipoles $C_{JLM}$ 
\begin{equation} 
\int_{0}^{\infty} {\rm d}q\,
C_{JLM}(q) =0 \, .
\end{equation}
As for the current, this condition is fulfilled when the CE is 
satisfied.


\section{Prescriptions for restoring the continuity equation}


The equations derived in the previous section assume that current
conservation is verified. Often, the nuclear current is not conserved
in electron scattering calculations, where a known density operator 
$\rho({\bf r})$, obtained as the sum of one-body single-nucleon 
densities, and a current operator ${\bf J}^{\rm NC}$, such as
\begin{equation}
\nabla \cdot {\bf J}^{(\rm NC)}({\bf r}) \, \not= \, -i\, 
\left[ H,\rho({\bf r}) \right] \, ,
\end{equation}
are considered.

Here we use the ICE to restore the CE. Starting with the
non-conserved current ${\bf J}^{(\rm NC)}$ and the charge density, 
we calculate the auxiliary current ${\bf C}^{(\rm NC)}$ as given by 
Eq.~(\ref{e8m}) and define new currents according to Eq.~(\ref{ICE00}). 
In particular we consider the values $q_0=0$ and $\infty$ for which 
$q_0{\bf J}({\bf q}_0)=0$ and we obtain
\begin{eqnarray}
\label{e35a}
{\bf J}^{(0)}({\bf q}) & = & 
-\frac{1}{q}\int_0^q {\rm d}q'\,{\bf C}^{(\rm NC)}({\bf q'}) \\
\label{e35b}
{\bf J}^{(\infty)}({\bf q}) &  = &
\frac{1}{q}\int_q^\infty {\rm d}q'\,{\bf C}^{(\rm NC)}({\bf q'}) \, ,
\end{eqnarray}
where ${\bf q'}=(q'/q){\bf q}$. As discussed in subsec. \ref{momen}, the
current ${\bf J}^{(0)}({\bf q})$ is the one derived by Friar and Fallieros
\cite{Fri84} and afterwards used by Friar and Haxton \cite{Fri85} to
calculate the electron scattering form factors. 

The second current ${\bf J}^{(\infty)}({\bf q})$ constitutes a new 
possible prescription in order to restore the CE. Both currents
coincide only if the model current is conserved; 
if CE is not verified, both prescriptions are in general different, 
and one has
\begin{equation}
{\bf J}^{(\infty)}({\bf q})-{\bf J}^{(0)}({\bf q}) = 
\frac{1}{q}\int_0^\infty {\rm d}q' \, {\bf C}^{(\rm NC)}({\bf q'}) 
\, \ne \, 0 \, .
\end{equation}

The asymptotic behavior of the new currents can be determined from 
their definitions in Eqs.~(\ref{e35a})-(\ref{e35b}). 
It can be summarized in the 
following properties:
\begin{eqnarray}
\label{e38a}
q\rightarrow 0 :~~~~~ & 
~~{\bf J}^{(0)}({\bf q}) \sim  -{\bf C}^{(\rm NC)}(0)\, , & ~~~~~~~~
{\bf J}^{(\infty)}({\bf q})  =   O(1/q) \\
\label{e38b}
q\rightarrow \infty :~~~~~ & 
{\bf J}^{(0)}({\bf q})  =  O(1/q) \, , & ~~~~~~~~
{\bf J}^{(\infty)}({\bf q})  \sim  {\bf C}^{(\rm NC)}({\bf q})
\end{eqnarray}

The first condition is equivalent to the Siegert theorem. In fact, 
from the definition of ${\bf C}({\bf q})$ in Eq.~(\ref{e8m}) we have 
\begin{equation}
-{\bf C}^{(\rm NC)}(0)\, = \, 
\left[ H,(\nabla_{\bf q}\rho)(0) \right] \,=\, 
i[H,{\bf d}] \, ,
\end{equation}
where ${\bf d}$ is the electric dipole momentum of the system
\begin{equation}
{\bf d} = \int {\rm d}^3r \, {\bf r} \,\rho({\bf r}) \, .
\end{equation}
Therefore the current ${\bf J}^{(0)}$ verifies the Siegert theorem because
it equals the time derivative of the electric dipole momentum of the 
system. 

The second current, ${\bf J}^{(\infty)}({\bf q})$, however, does not
satisfy this theorem. In fact, from its definition, we have
\begin{equation}
\lim_{q\rightarrow0}\,q{\bf J}^{(\infty)}({\bf q}) \,= \,
\int_0^\infty {\rm d}q\,{\bf C}^{(\rm NC)}({\bf q}) \, \ne \, 0
\end{equation}
Then, as established in (\ref{e38a}), this current diverges in the origin
as $O(1/q)$ if the original current is not conserved.
Therefore it should be discarded as a physical current for low
momentum transfer. 

On the other hand, a similar non-physical behavior is shown by the
Friar and Fallieros current ${\bf J}^{(0)}$ for large momentum 
transfer, because
\begin{equation}
\lim_{q\rightarrow\infty} \,
q{\bf J}^{(0)}({\bf q})\, =\,  
-\int_0^\infty {\rm d}q\,{\bf C}^{(\rm NC)}({\bf q})\, \ne \, 0 \, .
\end{equation}
This current goes as $O(1/q)$ for large $q$.

Instead, the current ${\bf J}^{(\infty)}$ works well in the large $q$
limit. From Eqs.~(\ref{e35a})-(\ref{e35b}) we can write
\begin{equation}
{\bf J}^{(\infty)}(q)\, = \,
\mu \int_0^\mu {\rm d}\mu' \,
\frac{1}{\mu'{}^2}\,{\bf C}^{(\rm NC)}({\bf q}') \, 
\stackrel{q\rightarrow\infty}{\longrightarrow} \,
{\bf C}^{(\rm NC)}({\bf q}) \, ,
\end{equation}
where the change of variables $\mu=1/q$, $\mu'=1/q'$ has been used.
Then the behavior of ${\bf J}^{(\infty)}$ for $q\rightarrow \infty$ is 
the same as that of the auxiliary current ${\bf C}^{(\rm NC)}$, which 
shows the adequate one because it is built
from the physical charge and current density operators
and its definition does not include integrations, but just 
derivatives and products. Its good asymptotic properties for large 
momentum transfer made this current to be {\em a priori}\/ useful in this
regime in electron scattering calculations, without the handicap of 
the pathological behavior shown by ${\bf J}^{(0)}$. The existence of such
a current was suggested in Ref. \cite{Fri85}, but, to the best of our
knowledge, this hypothesis has not been proved up to now.

In order to have a better understanding of the properties of these two 
currents, we can also obtain the asymptotic behavior in coordinate 
space. A straightforward calculation, similar to that developed for 
momentum space, permits to write:
\begin{eqnarray}
r\longrightarrow 0 :~~~~~ & 
{\bf J}^{(0)}({\bf r})  =  O(1/r^2) \, , &~~~~~
{\bf J}^{(\infty)}({\bf r})  \sim  {\bf J}^{(\rm NC)}({\bf r})
\\
r\longrightarrow \infty :~~~~~ & ~
{\bf J}^{(0)}({\bf r})  \sim  -{\bf C}^{(\rm NC)}({\bf r}) \, , &~~~~~
{\bf J}^{(\infty)}({\bf r})  =  O(1/r^2)
\end{eqnarray}
As we can see from these equations, both currents have also the opposite
roles. The current ${\bf J}^{(0)}$ is well behaved at large distances, 
but it diverges near the origin. On the other hand, for $r\rightarrow
0$, the current ${\bf J}^{(\infty)}$ goes as ${\bf J}^{(\rm NC)}$, which is
supposed to be well behaved at short distances, but it does not reach
zero fast enough for $r\rightarrow \infty$.

Finally, we summarize the asymptotic properties of the multipoles
$T^M_{JL}(q)$. By defining the multipoles of the two currents ${\bf
J}^{(0)}$ and ${\bf J}^{(\infty)}$ as
\begin{eqnarray}
\label{T0}
T^{(0)}_{JLM}(q) & = &
 -\frac{1}{q}\int_{0}^{q} {\rm d}q'\, C^{(\rm NC)}_{JLM}(q') \, , \\
\label{Tinf}
T^{(\infty)}_{JLM}(q) & = &
 \frac{1}{q}\int_{q}^{\infty} {\rm d}q' \, C^{(\rm NC)}_{JLM}(q') \, ,
\end{eqnarray}
respectively, we find the following asymptotic behavior
for these two sets of multipoles:
\begin{eqnarray}
q\rightarrow 0:~~~~~ &  ~~~~~~
T^{(0)}_{JLM}(q)  \sim  -C^{(\rm NC)}_{JLM}(q) \, , &~~~~~~~
T^{(\infty)}_{JLM}(q)  =  O(1/q)
\label{e49}\\
q\rightarrow \infty:~~~~~ &
T^{(0)}_{JLM}(q)  =  O(1/q) \, , &~~~~~~~
T^{(\infty)}_{JLM}(q)  \sim   C^{(\rm NC)}_{JLM}(q) \, .
\label{e50}
\end{eqnarray}
These properties are similar to the ones verified by the currents 
in momentum space given by Eqs.~(\ref{e38a})-(\ref{e38b}). 

The multipoles $T^{(0)}$ are well behaved for low momentum transfer, 
where the Siegert theorem applies, but fail for high momentum transfer.
As a consequence, the transition matrix elements corresponding to the 
current ${\bf J}^{(0)}$ do not go to zero fast enough as $q$ goes to
$\infty$, because the bound nuclear wave functions have an 
exponential-like behavior at large $q$ values. This is the reason of 
the pathology observed in this regime in the electric multipoles 
computed with the current developed by 
Friar and Fallieros \cite{Fri85,Cap97}. 

On the other hand, the multipoles $T^{(\infty)}$ do not work well near 
the origin (they diverge as $1/q$), but are well behaved for high 
momentum transfer. Hence they can safely be used in this regime 
in model calculations. 


\section{A simple model calculation}

In order to illustrate the results quoted in the last section, we have
used a very simple model to analyze different transitions in $^{16}$O 
and $^{39}$K and which was previously considered in Ref. \cite{Ama96}.
In this model, the nuclear structure is described by means of a
single-particle Hamiltonian of the form:
\[
H \, = \, -\frac{\hbar^2}{2m} \nabla ^2 \,+\, V_0 \,+ 
\,\frac{\hbar^2}{2m} \frac{r^2}{b^4}\, +
\,V_{\rm LS}\, {\bf l} \cdot {\bf s} \, .
\]
The eigenfunctions, $R_{nl}$, are harmonic oscillator functions and
the corresponding eigenvalues are
\[
\displaystyle
E_{nlj} \, = \, \frac{\hbar ^2}{mb^2} 
\left( 2n+l-\frac{1}{2} \right) \, +  \,V_0 \, +  \,
\frac{1}{2} \, V_{\rm LS} \, 
\left[ j(j+1)-l(l+1)-\frac{3}{4} \right] \, .
\]

The nuclear charge density operator is taken to be the usual one,
\[
\rho({\bf r})\, = \, \displaystyle \sum_{k=1}^A \, 
 \frac{1+\tau_3^k}{2} \, 
\delta({\bf r}-{\bf r}_k) \, ,
\]
while in the nuclear current density operator we have included the 
well-known convection and spin-magnetization one body terms,
\begin{eqnarray}
\label{convection}
{\bf J}^{\rm C}({\bf r}) & = & \displaystyle 
\displaystyle
\sum_{k=1}^{A} \frac{1}{2M_k} \frac{1}{i} \, \frac{1+\tau_3^k}{2} \, 
\left[ \delta({\bf r}-{\bf r}_k) \nabla_{{\bf r}_k} \, + \, 
\nabla_{{\bf r}_k} \, \delta({\bf r}-{\bf r}_k) \right] \, , \\
\label{magnetization}
{\bf J}^ {\rm M}({\bf r}) & = & \displaystyle \sum_{k=1}^{A}\,
\left(\mu_{\rm P}\frac{1+\tau_3^k}2+
      \mu_{\rm N}\frac{1-\tau_3^k}2 \right)
\nabla \times \left[ \delta({\bf r}-{\bf r}_k)
{\mbox{\boldmath $\sigma$}}^k \right] \, ,
\end{eqnarray}
as well as the so-called spin-orbit current,
\begin{equation}
\label{spin-orbit}
\displaystyle
{\bf J}^{\rm LS}({\bf r})\, =\, \frac{1}{2} \, V_{\rm LS} \,
\displaystyle \sum_{k=1}^A \, \frac{1+\tau_3^k}{2} \, 
\delta({\bf r}-{\bf r}_k) \,
{\mbox{\boldmath $\sigma$}}^k \times {\bf r}_k \, .
\end{equation}
In the previous equations $M_k$ labels the mass of the $k$--nucleon, 
${\bf S}^k={\mbox{\boldmath $\sigma$}}^k/2$ is its spin and
$\tau_3^k=1$ or $-1$ according this nucleon being proton or
neutron, respectively. Finally, $\mu_{\rm P}$ ($\mu_{\rm N}$) is
the proton (neutron) magnetic moment.

The model built in this way satisfies the CE. Within it we calculate 
the electric multipoles corresponding to the current
\begin{equation}
{\bf J}^{(\rm PD)} ({\bf r}) \, = \, {\bf J}^ {\rm C}({\bf r}) \, + \,
{\bf J}^ {\rm M}({\bf r}) \, + \, {\bf J}^ {\rm LS}({\bf r})
\end{equation}
for some transitions in $^{16}$O and $^{39}$K and the results 
are considered as ``pseudo-data''.
The parameters $V_0$, $V_{\rm LS}$ and $b$ used in the calculations 
are shown in Table~1. They were fixed in order to reproduce the 
energies of the single-particle states around the Fermi level in the
two double closed-shell nuclei ($^{16}$O and $^{40}$Ca) of interest.  

In what follows we discuss the point relative to the restoration of
the CE by using the currents ${\bf J}^{(0)}$ and 
${\bf J}^{(\infty)}$ defined 
in Eqs. (\ref{e35a}) and (\ref{e35b}), respectively. 
We have simulated the usual situation of a
model not verifying the CE by eliminating the spin-orbit current 
from the nuclear current operator:
\begin{equation}
{\bf J}^{(\rm NC)} ({\bf r}) \, = \, {\bf J}^ {\rm C}({\bf r}) \, + \,
{\bf J}^ {\rm M}({\bf r}) 
\end{equation}
We are interested in analyzing the
goodness of the approach sketched in the previous section in
retrieving the ``true'' results. 

We focus our attention on the two following electric transitions: 
$(0^+ \rightarrow 2s_{1/2} 1p^{-1}_{1/2})_{1^-}$ in $^{16}$O and 
$(1d^{-1}_{3/2} \rightarrow 2s^{-1}_{1/2})_{2^+}$ in $^{39}$K. 
In particular, we study the electric multipoles
\begin{equation}
\label{multe}
t^{(\kappa)}_{{\rm E}J}(q) \, = \, 
\sqrt{\frac{J +1}{2J+1}}
\langle J_{\rm f} \parallel i \, T^{(\kappa)}_{J J -1}(q) 
\parallel J_{\rm i} \rangle \, 
-\sqrt{\frac{J}{2J+1}}
\langle J_{\rm f} \parallel i \, T^{(\kappa)}_{J J +1}(q) 
\parallel J_{\rm i} \rangle \ , 
\end{equation}
where the multipole operators $T_{J J \pm 1}$ are given by 
Eq. (\ref{multipoles}) and where $(\kappa)$ stands for (PD), (0),
($\infty$) and (NC). 

In Fig. 1  we compare the multipoles
$|t^{(\kappa)}_{EJ}(q)|$ for $(\kappa)=(0)$ (full curves) and
($\infty$) (dashed curves) with the results of the calculation
done with the full model, $(\kappa)=({\rm PD})$ 
(dashed-dotted curves). The asymptotic
behavior discussed in the previous section is now apparent. The
calculations performed with ${\bf J}^{(0)}$, that is the Siegert's
current of Friar and Fallieros \cite{Fri84}, are right at low $q$, but
differs notably from the ``exact values'' in the high $q$ region,  
in agreement with the findings of Refs. \cite{Fri85,Cap97}. 
On the contrary,
the multipoles corresponding to ${\bf J}^{(\infty)}$ are wrong for small
momentum transfer, providing the correct behavior for large $q$. 

In order to have a better idea of the goodness of both calculations, we
show in Fig. 2 (upper panels (a,b) ) the quantity
\begin{equation}
\label{Delta}
\Delta^{(\kappa)}(q) \, \equiv \, 
|t^{(\rm PD)}_{EJ}(q)| - |t^{(\kappa)}_{EJ}(q)|
\end{equation}
for $(\kappa)=(0)$ (solid curves), $(\infty)$ (dashed curves) and (NC)
(dotted curves).

As we can see, this last calculation provides the better result except
at very low $q$, where we know it violates the Siegert theorem. 
However, for $q>1$ fm$^{-1}$ is closer to the ``pseudo-data'' than the
other two calculations. 

It is worth to note the situation of the
Siegert's current (solid curves). Although this current provides the right
behavior of the electric multipoles
at very low momentum transfer, they show a considerable
disagreement with the ``pseudo data'' 
for $q$ above 1 fm$^{-1}$, where the new
current we have obtained for $(\kappa)=(\infty)$ (dashed curves) gives
rise to more accurate results.

In order to go deeper in the analysis, we have done
new calculations without considering the magnetization current
(see lower panels (c,d)). As we know,
this current is not affected by the CE and by ignoring it we can test
the importance of this piece of the current in the results. However, we
can see that the situation does not change too much and similar
comments to those made above can be stated for these new calculations.
 
\section{Conclusions}

In this work we have developed a method to obtain new forms of the
nuclear electromagnetic current. The approach is based on the integral
form of the continuity equation and produces, as a particular case,
the Siegert's current developed by Friar and Fallieros. 
As in this case, the new currents can be used to restore
current conservation in those model calculations in which the
continuity equation is not fulfilled. Besides, our procedure permit to
understand in an easy way the asymptotic behavior shown by the
currents.

We have illustrated the method by means of a simple nuclear model
based on a harmonic oscillator potential which includes a spin-orbit
term. The results obtained show that, at least in the cases studied,
the multipoles calculated by using two of the new currents do not 
produce better results than those found with a model in which current 
is not conserved. This puts some doubts concerning the procedures of
``restoring the continuity equation''.

This situation is similar to the one found in relativistic
calculations of quasielastic electron scattering by nuclei
\cite{deF83} where no tractable approach to treating the off-shell
dependence rigorously exists. This makes inevitable the {\it ad hoc}\/
modifications of the currents in order to recover conservation, but    
it is not possible to decide which one of the different prescriptions is
the better.

In any case, our approach permits other possibilities to 
define new currents which deserve a more careful
analysis. Work in this direction is in progress.

\acknowledgements

This work has been supported by the DGICYT (Spain), under contract
PB95-1204, and by the Junta de Andaluc\'{\i}a (Spain).

\narrowtext
\begin{table}
\caption{Parameters of the potential used in the toy model
considered to discuss the violation of the CE (see text).}
\begin{tabular}{clll}
Nucleus   & $V_0$~[MeV] & $b$~[fm] & $V_{\rm LS}$~[MeV] \\
\tableline
\rule{0cm}{.4cm}
$^{16}$O  & -53.6       & 1.67     & -4.20 \\
$^{40}$Ca & -55.7       & 1.80     & -1.90 \\ 
\end{tabular}
\end{table}

\figure{
FIG.~1. Absolute value of the multipoles $t_{EJ}$ given by
Eq. (\protect\ref{multe}) as a function of the momentum transfer
$q$, for the two transition considered in this work. Dot-dashed 
curves correspond to the ``exact'' model. Full and
dashed curves have been obtained for the currents ${\bf
J}^{(\lambda_0)}$ for $\lambda_0=0$ and $\infty$, respectively.
\label{formf}
}

\figure{
FIG.~2. Differences $\Delta$ given by Eq. (\protect\ref{Delta}) for the
two transitions considered in this work. 
Full and
dashed curves have been obtained for the currents ${\bf
J}^{(\lambda_0)}$ for $\lambda_0=0$ and $\infty$, respectively. Dotted
curves correspond to the model in which the current is not conserved
(that is by ignoring the spin-orbit current
(\protect\ref{spin-orbit})). Upper panels (a,b) 
include the magnetization current
(\protect\ref{magnetization}) while this current has been not
included in the calculations plotted in the lower panels (c,d).
\label{differ}
}


\begin{references}

\bibitem{Sie37} A.J.F. Siegert, Phys. Rev. {\bf 52}, 787 (1937)

\bibitem{Ama96} J.E. Amaro, B. Ameziane and A.M. Lallena, 
                Phys. Rev. C {\bf 53}, 1430 (1996)

\bibitem{Cap97} E.C. Caparelli and E.J.V. de Passos,
                J. Phys. G: Nucl. Part. Phys. {\bf 23} (1997) 91

\bibitem{Hor97} Y. Horikawa and Y. Tanaka,
                Phys. Lett. B 409 (1997) 1

\bibitem{Nau97} H.W.L. Naus, 
                Phys. Rev. C 55 (1997) 1580;
                Nucl. Phys. A 628 (1998) 275

\bibitem{deF83} T. deForest, Nucl. Phys. {\bf A392}, 232 (1983);
                J.J. Kelly, Adv. Nucl. Phys. {\bf 23}, 75 (1996) 

\bibitem{Cab93} J.A. Caballero, T.W. Donnelly and G.I. Poulis, 
                Nucl. Phys. {\bf A555}, 709 (1993)

\bibitem{Fri82} J.L.Friar and S. Fallieros, 
                Phys. Lett. B {\bf 114},  403 (1982)

\bibitem{Fri84} J.L.Friar and S. Fallieros, 
                Phys. Rev. C {\bf 29}, 1645 (1984)

\bibitem{Fri85} J.L.Friar and W.C. Haxton, 
                Phys. Rev. C {\bf 31}, 2027 (1985)

\bibitem{deF66} T. deForest and J.D. Walecka,
                Adv. Phys. {\bf 15}, 57 (1966)

\end{references}
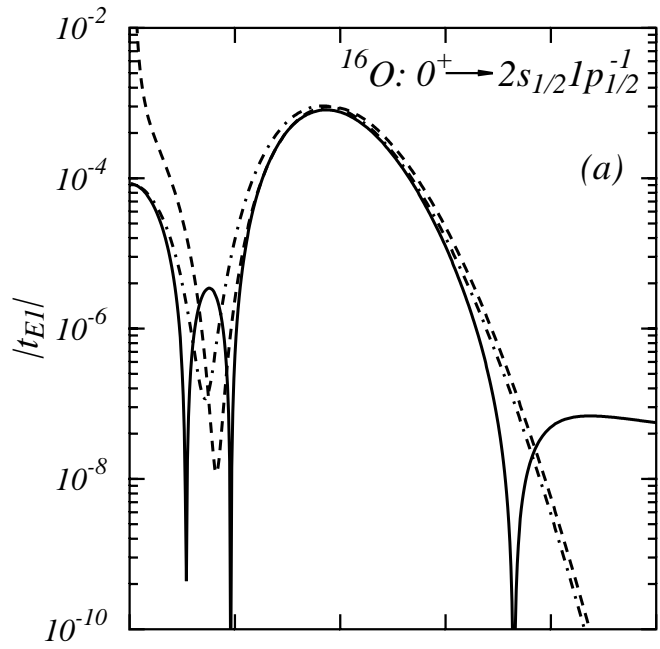
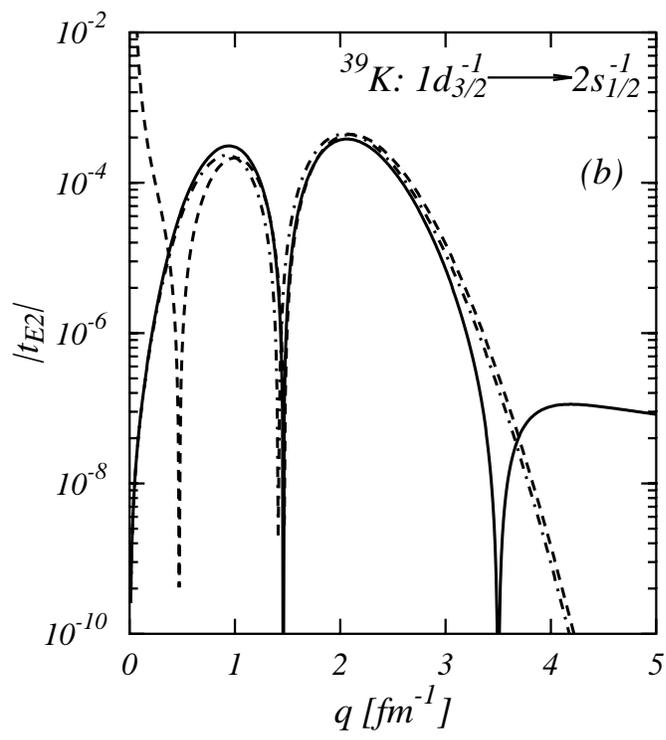
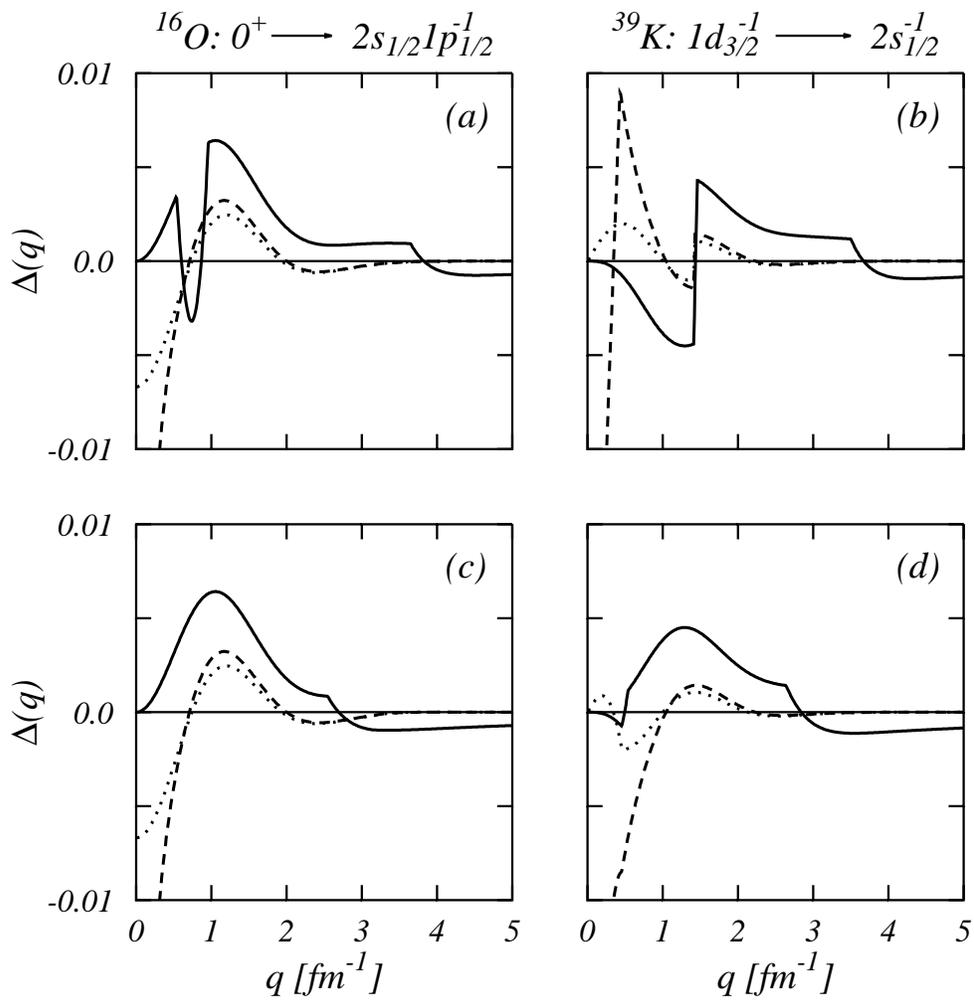
\end{document}